# Ferroelectricity in Atomically Thin Metallic TaNiTe$_5$ with Ultrahigh Carrier Density

Zhihua Liu[1], Shichong Song[1], Xunqing Yin[1], Chenhang Xu[1], Feng Liu[1], Guohua Wang[1], Peng Chen[1], Shengwei Jiang[1], Chunqiang Xu[2], Xiaofeng Xu[3], Weidong Luo[1], and Dong Qian[1, 4, 5*]

[1]State Key Laboratory of Micro-nano Engineering Science, Key Laboratory of Artificial Structures and Quantum Control (Ministry of Education), School of Physics and Astronomy, Shanghai Jiao Tong University, Shanghai 200240, China

[2]School of Physical Science and Technology, Ningbo University, Ningbo 315211, China

[3]School of Physics, Zhejiang University of Technology, Hangzhou 310023, China

[4]Tsung-Dao Lee Institute, Shanghai Jiao Tong University, Shanghai 200240, China

[5]Collaborative Innovation Center of Advanced Microstructures, Nanjing 210093, China

## Abstract

Ferroelectric metals, characterized by the coexistence of ferroelectricity and metallic conductivity, present a fundamental challenge due to the screening effect of free charge carriers on the long-range electric dipole order. Existing strategies to circumvent this obstacle include employing two-dimensional (2D) crystals, where reduced dimensionality and low carrier densities suppress screening, or designing materials of van der Waals (vdW) superlattice with spatially separated and decoupled conductive and nearly insulating ferroelectric layers. Here, we report an alternative paradigm in TaNiTe$_5$, where an ultrahigh carrier density coexists with an out-of-plane ferroelectric order within the same surface monolayer. Using piezoresponse force microscopy (PFM), we observed robust ferroelectric behavior in TaNiTe$_5$ down to single-unit-cell thickness (~1.3 nm) at room temperature. Scanning transmission electron microscopy (STEM) gives structural evidence that the ferroelectricity might originate from the vertical displacement of outmost Te atoms on the surface, breaking the inversion symmetry. Concurrently, electrical transport measurements reveal a metallic state with a carrier density on the order of $10^{15}$ cm$^{-2}$ (or $10^{22}$ cm$^{-3}$) -- comparable to that of Copper (Cu). Our findings establish a unique platform for exploring the interplay between ferroelectricity and an ultrahigh density of mobile carriers in the 2D limit.

*e-mail: dqian@sjtu.edu.cn

Ferroelectric metals represent a paradoxical class of materials where switchable electric polarization coexists with metallic conductivity, a phenomenon long considered mutually exclusive due to charge screening effects. Since the concept was first proposed in 1965 by Anderson and Blount[1], several intrinsic ferroelectric metals have been proposed[2-4]. However, experimental demonstrations of intrinsic ferroelectric metals have remained rare. Polar metals, such as $LiOsO_3$[5], $NdNiO_3$ film[6] and SiP monolayer[7] show metallic yet non-switchable polar states. In 2D materials, metallic and switchable ferroelectric state was observed in multilayer $WTe_2$[8, 9], and bilayer $T_d$-$MoTe_2$[10]. In those systems, the coexistence of metallic state and ferroelectric state is benefited from the reduced charge screening effects due to low-dimensionality and the low carrier density[8, 10, 11-13] which is on the order of $10^{13}$ cm$^{-2}$ (or $10^{19}$ cm$^{-3}$). For a traditional metal, such as Cu, the carrier density is ~ $10^{22}$ cm$^{-3}$ (or ~ $10^{15}$ cm$^{-2}$ for 1-nm thick Cu). Recently, coexistence of ferroelectricity and metallicity were also observed in weakly coupled $(SnSe)_{1.6}(NbSe_2)$ superlattice vdW crystal[14]. In this system, the overall carrier density is ~ $10^{21}$ cm$^{-3}$. However, the carriers are dominated by the non-ferroelectric $NbSe_2$ layers, while the ferroelectricity only exists in weakly conducting SnSe layers with very low carrier density. It is still an open question that whether robust ferroelectricity can exist in atomically thin and metallic system with carrier density comparable to traditional metal.

$TaNiTe_5$, a topological semimetal[15,16] with a bulk carrier density of ~ $10^{21}$ - $10^{22}$ cm$^{-3}$ [17-19] becomes a potential candidate for atomically thin ferroelectric metal with ultrahigh carrier density. Despite possessing a centrosymmetric bulk crystal structure, ferroelectric-like polarization was observed in our previous study on the surface of bulk $TaNiTe_5$[20]. If the polarization is limited on the surface, ferroelectricity could preserve during thickness reduction to few layers, overcoming the depolarization fields that typically destabilize ultrathin ferroelectrics. With complicated band structures[15,16,20] and high bulk carrier density[17-19], we expected the carrier density will hardly change in $TaNiTe_5$ few layers. Therefore, very likely, the atomically thin $TaNiTe_5$ could simultaneously exhibit ferroelectricity and ultrahigh carrier density within the same spatial region.

In this work, by combining PFM, electrical transport, and STEM measurements, we experimentally demonstrate a ferroelectric metallic state with ultrahigh carrier density in atomically thin $TaNiTe_5$ down to a thickness of few unit cells (UCs) at room temperature. The total carrier density for a four-UC-thick sample is ~ $6.4 \times 10^{15}$ cm$^{-2}$ at room temperature, which is on the same order of magnitude as that of Cu.

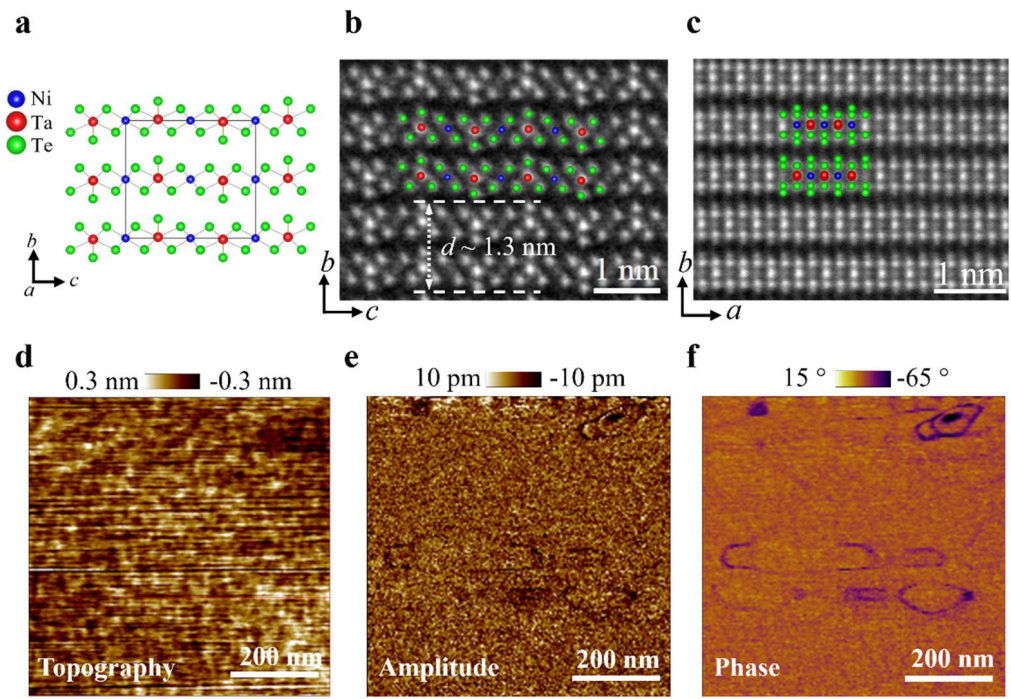

**Figure 1 | Crystal structure and out-of-plane ferroelectric domains of bulk TaNiTe₅. a,** Side view of the crystal structure. The unit cell is outlined in black lines. **b, c,** HAADF-STEM images acquired with incident electron beam along the [100] and [001] axes, respectively. Atomic structures from **a** were overlaid. **d,** PFM topography **e,** amplitude, **f,** phase images on natural cleaved (010) surface of TaNiTe₅.

Bulk TaNiTe$_5$ has a layered orthorhombic structure with the space group *Cmcm* (No. 63)[21]. As shown in Fig. 1(a), it comprises alternative TaTe$_3$ and NiTe$_2$ chains along the *c* axis. The vdW layers stack along the *b*-axis. One UC consists of two TaNiTe$_5$ monolayers. Lattice constants are $a = 0.3667\ nm$, $b = 1.3172\ nm$ and $c = 1.5142\ nm$[21]. Shown in Fig. 1b and 1c, the high-angle annular dark-field cross-sectional STEM (HAADF-STEM) images from our samples confirmed its crystal structure as Fig. 1(a) (more details are shown in Supplementary Note I). Bulk TaNiTe$_5$ adopts the centrosymmetric structure, which forbids intrinsic polarization due to the existence of inversion symmetry. However, PFM detects a pronounced response on the (010) surface of TaNiTe$_5$ in our previous report[20]. Detailed PFM measured hysteresis loops on the (010) surface of the bulk TaNiTe$_5$ crystal that we used to obtain few-layer TaNiTe$_5$ flakes are shown in Supplementary Note II. On the (010) surface of the bulk crystal, ferroelectric domains formed naturally are also observed by PFM. Our PFM is sensitive to the out-of-plane signals. Figures 1d-f present the PFM topography, amplitude, and phase images in the same region. No obvious patterns exist in topography image (Fig. 1d). In contrast, domain structures were observed both in amplitude (Fig. 1e) and phase images (Fig. 1f). The ring-like dark features exhibit opposite polarization in contrast to the rest of the region. Interestingly, the areas of the two opposite domains are unequal, suggesting that one domain may have lower energy. It should be noted that the detected phase difference between opposite domains deviates from the ideal 180°. We will later demonstrate that this deviation results from the relatively weak ferroelectric polarization and the characteristics of the PFM mapping technique.

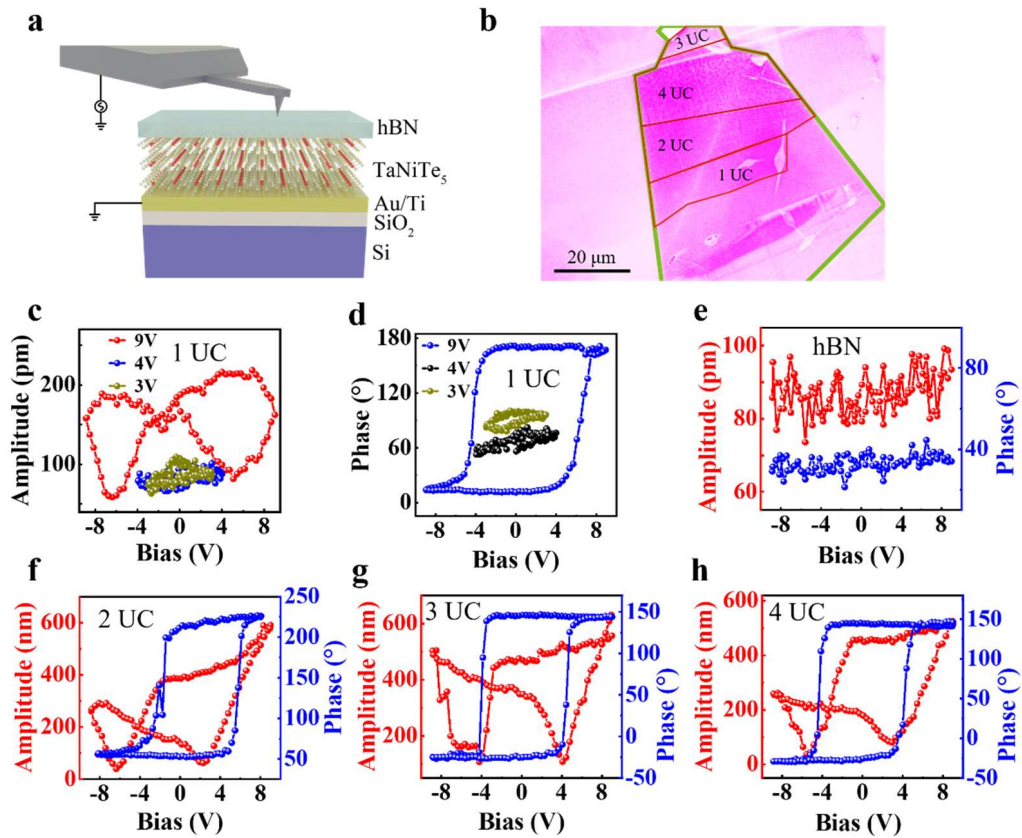

**Figure 2 | Out-of-plane ferroelectricity in few-layer TaNiTe$_5$. a,** Schematic illustration of the PFM experimental configuration on few-layer TaNiTe$_5$. **b,** Optical image of samples for PFM measurements. Green lines denote the region protected by hBN. Red lines denote the samples with different thicknesses under hBN. **c,** PFM amplitude and **d,** phase curves on 1-UC-thick TaNiTe$_5$, showing butterfly and hysteresis loops, respectively. **e,** PFM amplitude and phase curves on hBN flake. No ferroelectric signal was detected. PFM amplitude and phase curves on **f,** 2-UC, **g,** 3-UC, **h,** 4-UC TaNiTe$_5$ samples. Hysteresis loops in phase and butterfly loops in amplitude were observed.

The interlayer interaction in TaNiTe$_5$ is much stronger than graphene or WS$_2$, therefore we used Au assisted mechanical exfoliation method[22] to obtain few-layer-thick TaNiTe$_5$ (see Materials and Method in Supplementary Information). After being exfoliated in the glove box, the TaNiTe$_5$ flakes were encapsulated with ~ 5-nm-thick hBN, providing sufficient protection against environmental degradation while allowing penetration of a substantial electric field. The schematic experimental setup is shown in Fig. 2(a). Figure 2(b) presents the optical image of a TaNiTe$_5$ flake, where regions of different thicknesses coexist. The thicknesses of different regions were determined by AFM (see details in Supplementary Note III). In Figs. 2(c) and 2(d), 1-UC-thick (~ 1.3 nm) TaNiTe$_5$ exhibits ferroelectric signals including a butterfly-shaped amplitude loop and ~180° phase reversal with a clear threshold voltage. When scanning below the threshold voltage (e.g., between $\pm$ 4 V or $\pm$ 3 V), hysteresis loops disappear, as shown in Figs. 2c and 2d. The control experiment on pure 5 nm hBN shows negligible piezo response (Fig. 2e). Few-layer flakes without hBN encapsulation were also measured. As presented in Supplementary Note IV, the observation of clear ferroelectric hysteresis loops in samples without hBN encapsulation confirms that the ferroelectricity is intrinsic to TaNiTe$_5$ and not an artifact induced by hBN encapsulation. Ferroelectric hysteresis loops were observed also in 2-UC, 3-UC and 4-UC-thick TaNiTe$_5$ (Figs. 2(f)-(h)). Therefore, we can conclude that ferroelectricity persists in TaNiTe$_5$ from the few-layer regime down to the 1-UC. Since flakes with a half-UC thickness (~ 0.65 nm) cannot be reliably identified, we cannot determine whether ferroelectricity persists at this limit.

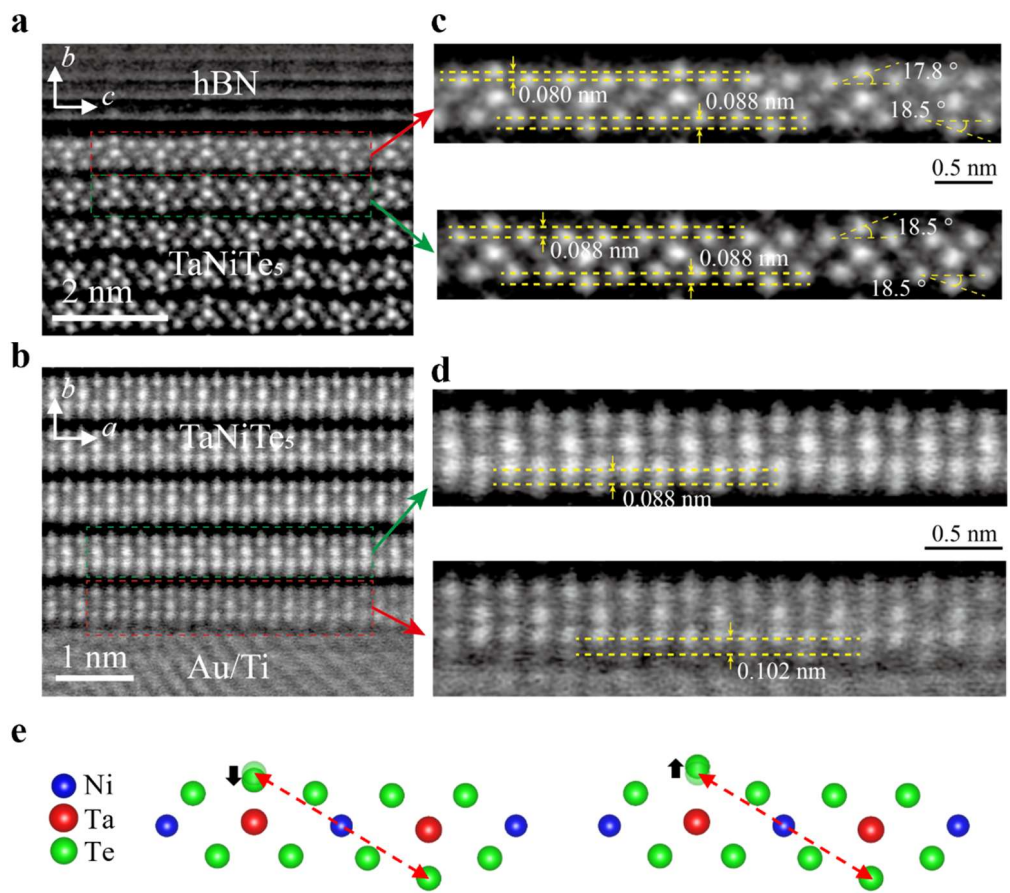

**Figure 3 | Inversion symmetry breaking on the surface monolayer of TaNiTe$_5$. a,** HAADF-STEM image near the top surface with incident electrons along the [100] axis. **b,** HAADF-STEM image near the bottom surface with incident electrons along the [001] axis. **c** and **d,** Zoom-in images from the regions defined by the red and green dashed rectangles in **a** and **b**. The yellow dashed lines mark the center positions of Te atoms. **e,** Schematic of the simplest atomic model illustrating the breaking of inversion symmetry in the surface TaNiTe$_5$ monolayer. Light green balls denote the original position of outmost Te atoms in the bulk TaNiTe$_5$ crystal. Black arrows indicate the displacement directions of the outmost Te atoms. Red dashed arrows indicate the inversion operation.

In our previous study, STM investigations on the ultrahigh-vacuum-cleaved surface of bulk TaNiTe$_5$ crystals revealed a complex surface relaxation that breaks the in-plane mirror symmetry[20]. In contrast, PFM measurements conducted were carried out in atmosphere and out-of-plane ferroelectric polarization was detected. Therefore, there might be no direct relation between the STM observed surface structure and the out-of-plane ferroelectric order. To further elucidate the microscopic origin of the ferroelectric behavior, we performed STEM measurements on the near-surface region of TaNiTe$_5$ flakes encapsulated by hBN -- the same samples used for the PFM measurements. HAADF-STEM images of the near-surface regions, specifically the top surface (hBN/TaNiTe$_5$) and bottom surface (TaNiTe$_5$/Au/Ti), are shown in Figs. 3a and 3b, respectively. Zoom-in views of the area marked by the red and green dashed rectangles in Figs. 3a and 3b are presented in Figs. 3c and 3d, respectively. The yellow dashed lines mark the center positions of Te atoms. On the top surface (Fig. 3c upper panel), the outmost Te atoms of the TaNiTe$_5$ monolayer displace out-of-plane by about - 9% (± 5%) relative to their positions in the bulk crystal structure (Fig. 3c bottom panel). In contrast, on the bottom surface (Fig. 3d bottom panel), the outmost Te atoms exhibit an out-of-plane displacement of about 16% (± 8%). Within experimental uncertainty, all other atoms retain positions consistent with the bulk structure. As summarized in Fig. 3e, this displacement of the outmost Te atoms breaks the inversion symmetry of the surface TaNiTe$_5$ monolayer, which may enable the emergence of surface ferroelectricity. Notably, the downward displacement of Te atoms (Fig. 3e) was observed in most surfaces, which might correlate with the unequal distribution of opposite domains shown in Fig. 1f.

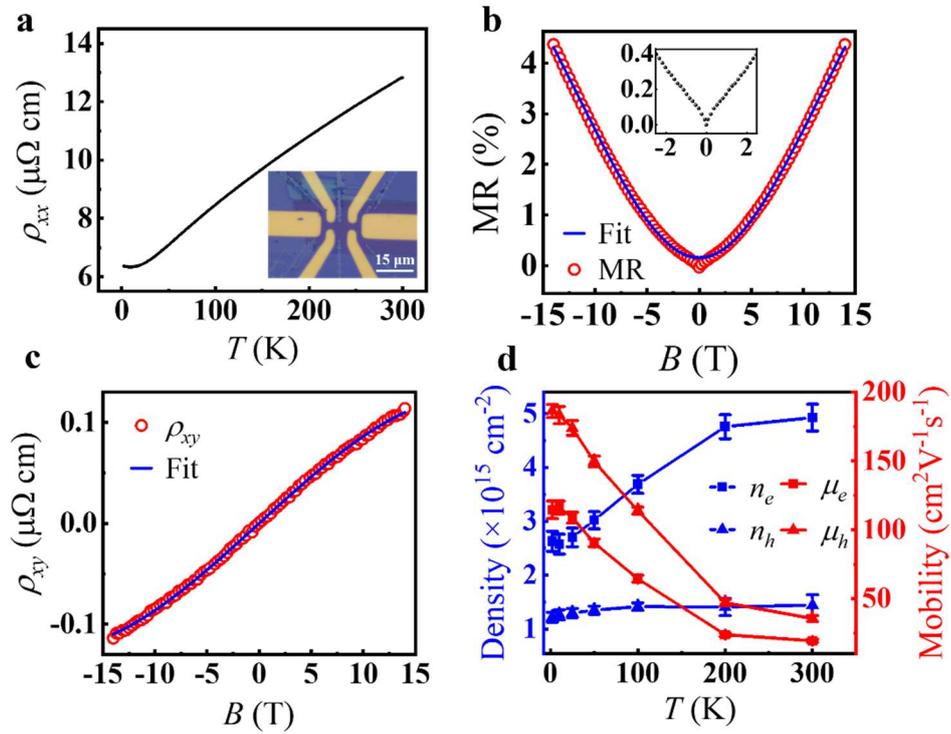

**Figure 4 | Electric transport properties of 4-UC TaNiTe$_5$. a,** Resistivity as a function of temperature. Current is applied along the *a*-axis and applied magnetic field B is along the *b*-axis.

Inset is the optical image of the device. **b**, MR at 2 K. Blue line is the fitting curve with two-band model. Inset is the MR in the low-field regime, showing the WAL effect. **c,** Hall resistivity at 2 K. Blue line is the fitting curve with two-band model. **d,** Carrier density (n) and mobility (μ) as a function of temperature.

We investigated the electrical transport properties of few-layer TaNiTe$_5$ samples. Due to the presence of cracks in the 1-UC to 3-UC samples obtained via Au-assisted mechanical exfoliation method, no conductive channel could be established between metal contacts on these samples. Therefore, the thinnest samples successfully measured in transport measurements were 4-UC thick. Fig. 4a shows the temperature-dependent resistivity ($\rho_{xx}$) of the 4-UC-thick flake, revealing metallic behavior with $\rho_{xx}$ decreasing from ~ 13 μΩ · cm at 300 K to ~ 6.2 μΩ · cm at 2 K. Figs. 4b and 4c present the magnetoresistance ($MR = \frac{\rho_{xx}(B) - \rho_{xx}(0)}{\rho_{xx}(0)}$) and Hall resistivity ($\rho_{xy}$) as a function of the magnetic field applied along *b*-axis at 2 K, respectively. The nonlinear behavior of $\rho_{xy}$ in Fig. 4c indicates the presence of multiple carriers. MR and $\rho_{xy}$ measured at different temperatures are provided in Supplementary Note V. Considering the semi-metallic band structure of TaNiTe$_5$[15,16], we analyzed the MR and $\rho_{xy}$ data using a two-band model described by the following equations:

$$\rho_{xx} = \frac{1}{e} \frac{(n_h \mu_h + n_e \mu_e) + (n_h \mu_e + n_e \mu_h) \mu_h \mu_e B^2}{(n_h \mu_h + n_e \mu_e)^2 + (n_h - n_e)^2 \mu_h^2 \mu_e^2 B^2}$$
$$\rho_{xy} = \frac{B}{e} \frac{(n_e \mu_e^2 - n_h \mu_h^2) - (n_h - n_e) \mu_h^2 \mu_e^2 B}{(n_h \mu_h + n_e \mu_e)^2 + (n_h - n_e)^2 \mu_h^2 \mu_e^2 B^2}$$

where $n_e$ ($n_h$) and $\mu_n$ ($\mu_h$) denote the electron (hole) concentrations and mobilities, respectively. The MR of 4-UC-thick TaNiTe$_5$ is well described by this two-band model, except in the low-field region. As shown in the inset of Fig. 4b, MR shows a sharp cusp, which is a hallmark of weak anti-localization (WAL). Fitting the MR(*B*) and $\rho_{xy}(B)$ curves at various temperature yields the temperature-dependent carrier densities and mobilities, shown in Fig. 4(d). At 2 K, the 4-UC sample exhibits comparable electron and hole density on the order of $10^{15}$ cm$^{-2}$ (or $10^{22}$ cm$^{-3}$), similar to bulk TaNiTe$_5$[17-19]. With increasing temperature, the electron density increases by about a factor of two, while the hole density increases only slightly. At 300 K, the carrier densities reach $n_e =$ ~ $4.9 \times 10^{15}\ cm^{-2}$; $n_h =$ ~ $1.4 \times 10^{15}\ cm^{-2}$, giving a total carrier density of ~ $6.3 \times 10^{15}\ cm^{-2}$. The carrier mobility is on the order of 100 cm$^{-2}$V$^{-1}$s$^{-1}$ at 2 K, smaller than that of bulk crystals[18,19], likely due to enhanced surface/interface scattering. Mobility decreases with increasing temperature. It should be noted that these electric transport measurements reflect the average behavior of the 4-UC TaNiTe5 flake. It remains possible that the carrier density in the top ferroelectric TaNiTe$_5$ monolayer differs significantly from that in the underlying layers. To address this, we carried out first-principles calculations, which shows that the displacement of Te atoms has a negligible effect on the band structure of TaNiTe$_5$ and, consequently, on the carrier density of the top TaNiTe$_5$ monolayer (see Supplementary Note VI).

In Table-I, we compared the total carry density of 4-UC-thick TaNiTe$_5$ with that of other metallic ferroelectric materials and Cu. Significantly, ferroelectricity persists in few-layer TaNiTe$_5$ despite its carrier density being comparable to that of Cu and about two orders of magnitude higher than that of other ferroelectric metals.

Table-I Total carrier density of ferroelectric metals

| Materials | Temperature | 3D carrier density | 1 nm carrier density | Curie Temperature |
|---|---|---|---|---|
| Cu[23] | 300 K | $8.5\times10^{22}$ cm$^{-3}$ | $8.5\times10^{15}$ cm$^{-2}$ | None ferroelectric |
| TaNiTe$_5$ (this work) | 300 K | $1.2\times10^{22}$ cm$^{-3}$ | $1.2\times10^{15}$ cm$^{-2}$ | >300 K |
| Few-layer WTe$_2$[24] | 1.8 K | $1\times10^{20}$ cm$^{-3}$ | $1\times10^{13}$ cm$^{-2}$ | - |
| Bulk WTe$_2$[25] | 300 K | $5\times10^{20}$ cm$^{-3}$ | $5\times10^{13}$ cm$^{-2}$ | >300 K |
| Few-layer WTe$_2$[8] | 7 K | $6.3\times10^{19}$ cm$^{-3}$ | $6.3\times10^{12}$ cm$^{-2}$ | >300 K |
| Bi$_2$O$_2$Se[26] | 300 K | $1.8\times10^{19}$ cm$^{-3}$ | $1.8\times10^{12}$ cm$^{-2}$ | >300 K |

Furthermore, we examined the stability of ferroelectric polarization under such high carrier density in Fig. 5. The applying a ± 9 V bias through the AFM tip, rectangular domains were successfully written and subsequently confirmed by PFM imaging (Figs. 5a-c). These ferroelectric domains maintained both phase contrast and morphological integrity. Both the phase contrast and domain shape remained nearly unchanged after 1 hour (Figs. 5(g)-(i)). The persistence of written ferroelectric domains for over 1 hour is comparable to other vdW ferroelectric materials, such as 1T'-MoS$_2$[27] and ReS$_2$[28]. Although the written patterns exhibit clear phase contrast, the phase difference between opposite domains deviates from the ideal 180°, a behavior usually observed in systems with relatively weak ferroelectric polarization[29-31]. In fact, as demonstrated in Supplementary Note VII, the phase difference between opposite domains can be recovered to 180°.

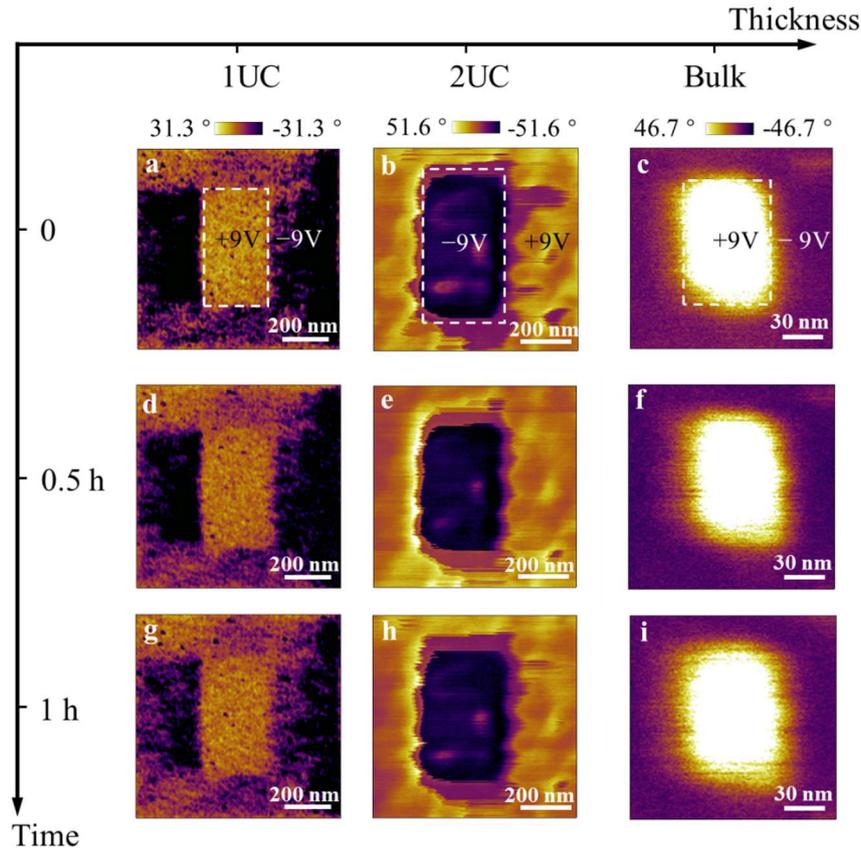

**Figure 5 | Domain patterning PFM phase images of hBN-encapsulated TaNiTe$_5$ with varying thicknesses. a,** PFM phase after writing for **b**, 1 UC, (f) 2 UC, and (j) bulk, sample was applied by ± 9 V bias. Phase evolution after 30 minutes for (c) 1 UC, (g) 2 UC, and (k) bulk. And after 1 hour for (d) 1 UC, (h) 2 UC, and (l) bulk.

In summary, our comprehensive PFM, electrical transport and STEM measurements demonstrate the coexistence of room-temperature out-of-plane ferroelectricity and metallic conductivity with ultrahigh carrier density in atomically thin TaNiTe$_5$. This unconventional ferroelectricity originates from the surface monolayer of TaNiTe$_5$ and is likely associated with the vertical displacement of outmost Te atoms. Our findings provide the first example that ferroelectricity and ultrahigh carrier density coexist within the same spatial region.

**Acknowledgements**

This work was supported by the National Key R&D Program of China, the National Natural Science Foundation of China.

**Author contributions**

D.Q. conceived the project and designed the experiments. Z.L., X.Y., and X. X. grew the single crystal. Z.L. fabricated the devices fabrication and performed PFM. Z.L. and C.X. performed STEM measurement. Z.L. and G.W. performed electrical transport measurement. S.S. and W.L. performed DFT calculations. Z.L., S.S., W.L., F.L., S.J. and D.Q. analyzed the data. Z.L. and D.Q. wrote the paper. All the authors discussed the results and commented on the manuscript.

**Competing financial interests**

The authors declare no competing financial interests.

**Correspondence and requests for materials** should be addressed to Dong Qian.